\documentclass[10pt]{iopart}
\usepackage{graphicx}
\newcommand {\be}{\begin{equation}}
\newcommand {\ee}{\end{equation}}

\begin{document}

\paper[Anomalous kinetics from 1D MCT]
{Anomalous kinetics and transport from 1D self--consistent 
mode--coupling theory}
\date{\today}

\author{Luca Delfini $^{1,2}$, Stefano Lepri $^{1}$,  
Roberto Livi $^{2,3}$  and Antonio Politi $^{1}$\
}
\address{$^{1}$ Istituto dei Sistemi Complessi, Consiglio Nazionale
delle Ricerche, via Madonna del Piano 10, I-50019 Sesto Fiorentino, Italy
}
\address{$^{2}$ Dipartimento di Fisica, Universit\`a di Firenze, 
via G. Sansone 1 I-50019, Sesto Fiorentino, Italy 
}
\address{$^{3}$ Sezione INFN, Unit\'a INFM and CSDC Firenze,
via G. Sansone 1 I-50019, Sesto Fiorentino, Italy } 

\ead{luca.delfini@fi.isc.cnr.it}

\begin{abstract}
We study the dynamics of long-wavelength fluctuations in one-dimensional 
(1D) many-particle systems as described by self-consistent mode-coupling theory.
The corresponding nonlinear integro-differential equations for the relevant 
correlators are solved analytically and checked numerically. In particular, we
find that the memory functions exhibit a power-law decay accompanied by
relatively fast oscillations. Furthermore, the scaling behaviour and,
correspondingly, the universality class depends on the order of the leading
nonlinear term. In the cubic case, both viscosity and thermal conductivity
diverge in the thermodynamic limit. In the quartic case, a faster decay of the
memory functions leads to a finite viscosity, while thermal conductivity
exhibits an even faster divergence. Finally, our analysis puts on a more firm
basis the previously conjectured connection between anomalous heat conductivity
and anomalous diffusion.

\end{abstract}

%\submitto{Journal of Statistical Mechanics: theory and experiment}

%\maketitle
\section{Introduction}

Numerical and theoretical studies of transport in reduced spatial dimensions (1
and 2D) have revealed that unexpected properties of matter may emerge with
respect to the standard features of 3D systems. This fact, combined with the
possibility of designing and manipulating matter at the nanoscale, unveals the
possibility  of exploring new physical phenomena and applications. Preliminary
indications about anomalous hydrodynamic behaviors were discovered in the
context of non--equilibrium statistical mechanics by the appearance of 
long--time tails in fluid models \cite{PR75,kirk}. In fact, there are many
instances showing that microscopic models of hydrodynamics can be affected by
serious pathologies. For the sake of an example, we mention the anomalous
scaling of Rayleigh and Brillouin peak widths in the hydrodynamic limit
detected in the 1D Lennard--Jones fluid \cite{Bishop2,sandri}. Another instance
is the divergence of viscosity in cellular automata fluids \cite{DLQ89} that may
however be due their peculiar features. 

On the one hand, the breakdown of phenomenological constitutive laws of
hydrodynamics, yielding ill-defined transport coefficients, could be viewed as
a drawback for the adopted models and theoretical approaches. On the other hand,
it is nowadays widely recognized that anomalous transport is a documented
effect rather than an accident due to the oversimplification of theoretical
models. Typical examples are single-filing systems, where particle diffusion
does not follow Fick's law~\cite{singlef}, and the enhancement of vibrational
energy transmission in quasi-$1D$ systems, such as polymers~\cite{morelli} or
individual carbon nanotubes~\cite{nanotube}. 

Within this general context, one of the issues that attracted a renewed 
interest in the last decade is the problem of anomalous heat conduction in 
low-dimensional many-particle systems \cite{LLP97,LLP98}. In this case, the
anomalous features amount to the divergence of the finite-size heat
conductivity $\kappa(L)\propto L^\alpha$ in the limit $L\to \infty$ and,
correspondingly, to a nonintegrable decay of equilibrium correlations of the
energy current (the Green-Kubo integrand),
$\langle J(t)J(0)\rangle \propto t^{-(1-\alpha)}$($0\le\alpha < 1$) 
for long times $t\to \infty$. Numerical studies ~\cite{rep}
and theoretical arguments \cite{NR02} indicate that anomalies should occur
generically in 1 and 2D, whenever momentum is conserved. In particular, a great
deal of studies have been devoted to the prediction of the exponent $\alpha$
for a twofold reason. On a basic ground, one is interested in identifying the
universality classes and the relevant symmetries possibly identifying them.
From a practical point of view, an accurate prediction of the scaling behaviour
allows determining heat conductivity in finite systems -- a crucial issue for
the comparison with experiments. 

Despite many efforts, the theoretical scenario is still controversial. The first
attempt of tackling the problem was based on arguments derived from the
Mode-Coupling Theory (MCT). This approach proved quite powerful for the estimate
of long-time tails in fluids \cite{PR75} and for describing the glass transition
\cite{mct}. In 1D models \cite{E91}, the MCT prediction was $\alpha = 2/5$ 
\cite{LLP98,L98}, very close to the numerical values \cite{rep}. This estimate
was later criticized as inconsistent in \cite{NR02}, where renormalization group
arguments were instead shown to yield $\alpha =1/3$. Nevertheless, the 2/5 value
has been later derived both from a kinetic-theory calculation for the quartic
Fermi--Pasta--Ulam (FPU-$\beta$) model \cite{P03} and from a solution of the MCT
by means of an {\it ad hoc} Ansatz \cite{li}. It was thereby conjectured
\cite{li} that 2/5 is found for a purely longitudinal dynamics, while a
crossover towards 1/3 is to be observed only in the presence  of a coupling to
transversal motion. Unfortunately, the accuracy of numerical simulations is
generally insufficient to disentangle the whole picture. Without pretending to
give a full account of the various numerical results (see \cite{chaos} for a
recent account), it is worth mentioning that, while for some models like the so
called hard--point gas \cite{hpg} the data is compatible with the prediction
$\alpha = 1/3$ \cite{denis}, some other cases show huge deviations. For example,
in the purely quartic FPU model, $\alpha$ is definitely larger than 1/3 and 
definitely closer to 2/5 \cite{LLP03}. The situation is even more controversial
in 2D, where MCT predicts a logarithmic divergence \cite{2d} that is not yet
confirmed by numerical simulations.

For all the above reasons, we decided to reconsider MCT to clarify
the origin of the reported discrepancies. Typically, MCT is implemented by 
imposing an ad-hoc Ansatz for the dependence of the hydrodynamic propagator and
of the memory kernel on the space and time variables. In this paper, following
~\cite{DLLP06}, we find an exact self-consistent solution of the MCT
equations, without any \textit{a priori} assumption on the form of the solution.
The resulting scaling behavour differs from that one proposed in \cite{li}. 
Moreover, the asymptotic behaviour turns out to depend on the order of the
leading nonlinearity in the interaction potential. Indeed, we find that cubic
and quartic nonlinearities are characterized by $\alpha =1/3$, and $1/2$,
respectively. Remarkably, an even more striking difference is exhibited by
viscosity which appears to diverge in the former case, while is finite in the
latter one.

The structure of the paper is as follows. In Section \ref{s:mct} we review the
mode--coupling equations and discuss their physical meaning. Section
\ref{s:cubic} is devoted to the solution when the leading nonlinearity is
cubic, while in Section \ref{s:quartic} we discuss how the
solutions change in the presence of a leading quartic nonlinearity. In Section
\ref{s:cur}, the knowledge of the mode--coupling solutions is exploited to
determine the value of the scaling exponent $\alpha$, by estimating the heat
current correlation function. Final remarks and future perspectives of this
approach are reported in the conclusions.

\section{MCT equations}
\label{s:mct}

The introduction of suitable stochastic equations ruling the dynamics of the
relevant variables is rather customary to describe the relaxation close to
equilibrium \cite{KT}. The idea consists in concentrating on the effective motion
of the ``slow'' observables, i.e. in changing level of description. The
general strategy amounts to projecting the original equations onto a suitable
subspace and eventually leads to linear non-Markovian equations. The memory
term determines the relaxation properties. Whenever a sharp separation of time
scales exist, the equations reduce to their Markovian limit.  

Let us consider the simplest one-dimensional version of the self-consistent MCT.
For definiteness, we consider a Hamiltonian chain of oscillators,
interacting through a generic nearest-neighbour coupling potential $V$, whose
Taylor expansion around equilibrium is written as
\begin{equation}
V(y) \;=\; {1\over 2}
 \, y^2 \,+\, {1\over 3}g_3 \, y^3 \,+\, {1\over 4}g_4 \, y^4 \, + \, \ldots
\label{fpu}
\end{equation}
The variable $y$ is a short-hand notation for the difference of nearest
neighbour displacement fields $(u_{i+1} - u_i)$, with the integer index
$i$ labelling the lattice sites. Besides the conservation of total energy,
such a kind of interaction implies also the conservation of total momentum
due to space-translation invariance.

The main observable we are interested in is the normalized correlator 
$G(q,t)= \langle Q^*(q,t)Q(q,0) \rangle/\langle
|Q(q)|^2 \rangle$, where $Q(q,t)$ is the Fourier transform of the 
displacement field $u_i(t)$. By assuming periodic boundary conditions for 
a chain made of $N$ sites, the wavenumber is given by $q=2\pi k/N$, 
with $-N/2+1 \leq k \leq N/2$. Notice also that
$G(q,t)=G(-q,t)$. We simplify the notation by setting to 
unity the particle mass, the lattice spacing and the bare sound velocity. 
The equations for the correlator $G(q,t)$ then read ~\cite{SS97,L98}
\begin{equation}
\nonumber
{\ddot G} (q,t) + 
\varepsilon \int_0^t \Gamma (q,t-s) {\dot G}(q,s) \, ds 
+ {\omega}^2(q) G(q,t)  
= 0
\label{mct}
\end{equation}
where the memory kernel $\Gamma(q,t)$ is proportional to $\langle
\mathcal{F}(q,t)\mathcal{F}(q,0) \rangle$ with $\mathcal{F}(q)$ being the
nonlinear part of the force between particles. Equations (\ref{mct}) must be
solved with the initial conditions $G(q,0)=1$  and $\dot G(q,0)=0$. 
Equations ~(\ref{mct}) are exact and they are derived within the well--known
Mori--Zwanzig projection approach \cite{KT}. 

The mode--coupling approach basically amounts to replacing the exact memory
function $\Gamma$ with an approximate one, where higher--orders correlators are
written in terms of the $G(q,t)$. This yield a closed system of nonlinear
integro--differential equations. For potentials like (\ref{fpu}) this has been
worked out in detail in references ~\cite{SS97,L98}. Both the coupling constant
$\varepsilon$ and the frequency $\omega(q)$ are temperature-dependent input
parameters, which should be computed independently by numerical simulations or
approximate analytical estimates \cite{SS97,L98}.  For the aims of the present
work, we may restrict ourselves to considering their bare values, obtained in
the harmonic approximation. In the adopted dimensionless units they read
$\varepsilon = {3g_3^2 k_BT / 2\pi}$ and  $\omega(q)=2 | \sin{q\over 2}|$. Of
course, the actual renormalized values are needed for a quantitative comparison
with specific models. 

To understand the physical interpretation
of equation (\ref{mct}), let us first assume that a Markovian approximation holds i.e. we
can replace the memory function by a Dirac delta. In the small wavenumber limit
$q\to 0$ they reduce to 
\begin{equation}
{\ddot G} (q,t) + \eta q^2 {\dot G}(q,t) +  c^2q^2 G(q,t)  \;=\; 0
\label{markov}
\end{equation}
where, for the sake of clarity, we have reintroduced the sound velocity $c$.
Notice that the validity of (\ref{markov}) rests on the fact that $\Gamma$
exhibits a fast decay and that its time integral is finite.
This equation describes the response of an elastic string at finite 
temperature as predicted by linear elasticity theory \cite{LL} for 
the macroscopic displacement field $u(x,t)$
\be
\ddot u - c^2\partial_x^2 u - \eta \partial_x^2 \dot u \;=\; 0\quad.
\ee
Here $\eta$ is a suitable viscosity coefficient and describes the internal
irreversible processes of the elastic string (i.e. the absorption of ``sound''
waves). We anticipate that one of our results is precisely that 
(generically) the above approximation \textit{does not hold} and that 
long--range memory effects must be retained \cite{L00}. 

\section{Cubic nonlinearities}
\label{s:cubic}

In the generic case in which $g_3$ in equation~(\ref{fpu}) is different from
zero, the lowest-order mode coupling approximation of the memory kernel turns
out to be \cite{SS97,L98} 
\begin{equation}
\Gamma(q,t)= \,\omega^{2}(q)
\,\frac{2 \pi}{N} \sum_{p+p'-q=0,\pm\pi}  \,G(p,t) G(p',t) \quad .
\label{mct2}
\end{equation}
Equation (\ref{mct}) along with (\ref{mct2}) are now a closed system that
has to be solved self-consistently. The integral 
terms contain products of the form $GG\dot G$ and we thus refer to 
this case as to the cubic nonlinearity.

\subsection{Analytical results}

Direct numerical simulations \cite{L98} indicate that for small wavenumbers
nonlinear and nonlocal losses in equation~(\ref{mct}) are small compared with the
oscillatory terms. This suggests splitting the $G$ dynamics into phase and
amplitude evolution,
\begin{equation}
G(q,t) \;=\; C(q,t )e^{i \omega (q)t} + c.c.
\label{g}
\end{equation}
As already shown in \cite{DLLP06}, this Ansatz allows reducing the dynamics to
a first order equation for $C(q,t)$ and, in turn, determining the scaling
behavior of the correlators.
Let us start by rearranging the expression for $\Gamma(q,t)$.
By inserting equation~(\ref{g}) into (\ref{mct2}), we obtain
\begin{eqnarray}
\Gamma (q,t) &\;=\;&  \omega^2(q) \int_{-\pi}^{\pi} dp \, \Big[  
C(p,t)C(q-p,t )e^{i[\omega (p)+\omega (q-p)]t} \\
&& +C(p,t)C^{\ast}(q-p,t )e^{i[\omega (p)-\omega (q-p)]t} \Big] \,+\, c.c.
\nonumber
\end{eqnarray}
where the sum in (\ref{mct}) has been replaced by an integral, since
we are interested in the thermodynamic limit $N=\infty$. For small
$q$-values, which are, by the way, responsible for the
asymptotic behavior, the above equation reduces to
\begin{eqnarray}
\Gamma (q,t) &\;\approx &\;  q^2 \int_{-\pi}^{\pi} dp \, \Big[
C(p,t)C(q-p,t)e^{2i\omega (p)t} \\
&& +C(p,t)C^{\ast}(q-p,t )e^{-i\omega'(p)qt} \Big] \,+\, c.c.
\nonumber
\end{eqnarray}
where $\omega '(p)= \frac{d\omega}{dp}$.
The first term in the r.h.s. and its complex conjugate are negligible because
of the rapidly varying phase. By further neglecting the dependence of the sound
velocity on $q$ (i.e. $\omega '(p)\approx 1$), one finds that $\Gamma(q,t)$ has
the following structure,
\begin{equation}
\Gamma (q,t) \; =\;   M(q,t) e^{iqt}\, +\, c.\,c.
\label{Gamma}
\end{equation}
where 
\begin{equation}
M (q,t) \equiv  q^2 \int_{-\pi}^{\pi} dp \, C^{\ast}(p,t)C(q-p,t) 
\label{emme}
\end{equation}
is the kernel in the new description.
Upon substituting equation~(\ref{Gamma}) and (\ref{g}) into (\ref{mct}), 
one obtains, in the slowly varying envelope approximation, $\dot{C} \ll qC $,
\begin{equation}
2 \dot C(q,t ) + 
\varepsilon \int_0^{t}\, dt ' \, M(q,t-t ') C(q,t ') \;=\; 0
\label{eqc}
\end{equation}
and a similar expression for $C^*$. 

Equation (\ref{eqc}) have been obtained after discarding the second order time
derivative of $C(q,t)$ as well as the integral term
proportional to $\dot C$, besides all rapidly rotating terms. The validity
of this approximation is related to the separation between the decay rate of
$C(q,t)$ and $\omega(q)$; its correctness will be cheked a posteriori, after
discussing the scaling behaviour of $C(q,t)$. Notice also that in this
approximation, Umklapp processes do not contribute: it is in fact well
known that they are negligible for long--wavelength phonons in 1D ~\cite{rep}. 

Having transformed the second order differential equation into a first
order one for $C(q,t)$, it is now possible to develop a scaling analysis
for the dependence of $C$ on $q$ and $t$, owing to the homogeneous structure
of the resulting equation (see also \cite{LB94}, where a similar equation was
investigated). More precisely, one can easily verify that the following
Ansatz holds, 
\begin{equation}
C(q,t) = g(\sqrt{\varepsilon}t q^{3/2}) \quad,\quad   
M(q,t) = q^{3}f(\sqrt{\varepsilon}t q^{3/2}) .
\label{gf3}
\end{equation}
This shows that the decay rate for the evolution of $C(q,t)$ is given by 
$q^{3/2} \sqrt{\varepsilon}$. Since the corresponding rate for the phase factor
is $q$, one can conclude that amplitude and phase dynamics are 
increasingly separated for $q \to 0$. High $q$-values ($q \approx 1$) are those
for which the slowly varying envelope approximation is less accurate. However,
if $\varepsilon$ is small enough, such modes are correctly described, too.
This has been checked in the numerical solution of equation~(\ref{mct}) 
(see the following subsection).

The functions $f$ and $g$ can be determined by substituting the 
previous expressions into (\ref{eqc}) and(\ref{emme}). Upon setting
$x=\sqrt{\varepsilon}t q^{3/2}$, one obtains the equation
\begin{eqnarray}
&&\frac{dg}{dx} = -\int_0^x dy \, f(x-y) g(y)
\label{gx}\\
&&f(x) = x^{-2/3}\int_{-\infty}^{+\infty} dy
g^\ast (\mid x^{2/3}-y\mid ^{3/2}) g(y^{3/2}) \quad,  
\label{fx}
\end{eqnarray}
where the integral in (\ref{emme}) has be extended to infinity.

The asymptotic behaviour for $x\to 0$ can be determined analytically.
Actually, one can neglect the dependence on $x$ in the integral appearing in
(\ref{fx}), thus obtaining
\begin{equation}
f(x) \;=\; \frac{a}{x^{2/3}}
\label{effe}
\end{equation}
where
\begin{equation}
a = \int_{-\infty}^{+\infty} dy |g|^2 (y^{3/2})  
\end{equation}
which implies that
\begin{equation}
M(q,t) = \frac{a}{\varepsilon^{1/3}} \frac{q^{2}}{t^{2/3}} 
\label{m}
\end{equation}
By replacing equation~(\ref{effe}) into (\ref{gx}) and introducing the
dummy variable $z=x-y$, one finds
\begin{equation}
\frac{dg}{dx} = -a \int_0^x dz \frac{g(x-z)}{z^{2/3}}  =
   - 3a g x^{1/3} -\frac{3}{2}a \frac{dg}{dx} x^{4/3} + \ldots
\end{equation}
In the small-$x$ limit, the last term in the r.h.s. is negligible with respect
to the term in the l.h.s. After solving the resulting differential equation one
obtains,
\begin{equation}
g(x) \;=\; \frac12 \exp \Big(-{9 a x^{4/3}\over 4} \Big) 
\label{smallx}
\end{equation}
where the multiplicative factor has been determined by imposing the
normalization condition $G(0) = 1$.

Under the approximation that expression (\ref{effe}) holds for every $x$, one
can exactly solve equation~(\ref{gx}) by Laplace transformation. As a result, one
obtains 
\begin{equation}
C(q,z) \;=\; \frac{iz^{1/3}}{iz^{4/3}+ a q^{2}}\quad.
\end{equation}
This expression is precisely the Laplace transform of the Mittag-Leffler
function $E_{\mu}(-(\lambda t)^{\mu})$~\cite{maina,adiff} for $\mu = 4/3$ and
$\lambda= (a q^2)^{3/4}$. 
The function $E_\mu$ interpolates between stretched 
exponential and power-law decays (see Appendix B in \cite{adiff}). 
The asymptotic form for small values of the function argument yields the same 
dependence for $C$ (see the first of equations~(\ref{gf3})).
This observation suggests that the effective evolution of 
fluctuations should be represented by the fractional differential equation
\be
{\partial ^\mu \over\partial t ^\mu} C(q,t) + \lambda^\mu  C(q,t) 
\;=\; 0 .
\label{fractional}
\ee
The case of interest here ($1<\mu\le 2$) corresponds to the so--called fractional 
oscillations \cite{maina}. It should be emphasized that in the present context,
memory arises as a genuine many-body effect and is not postulated 
{\it a priori}.

\subsection{Numerical analysis}

\begin{figure}
\begin{center}
\includegraphics[clip,width=7.5cm]{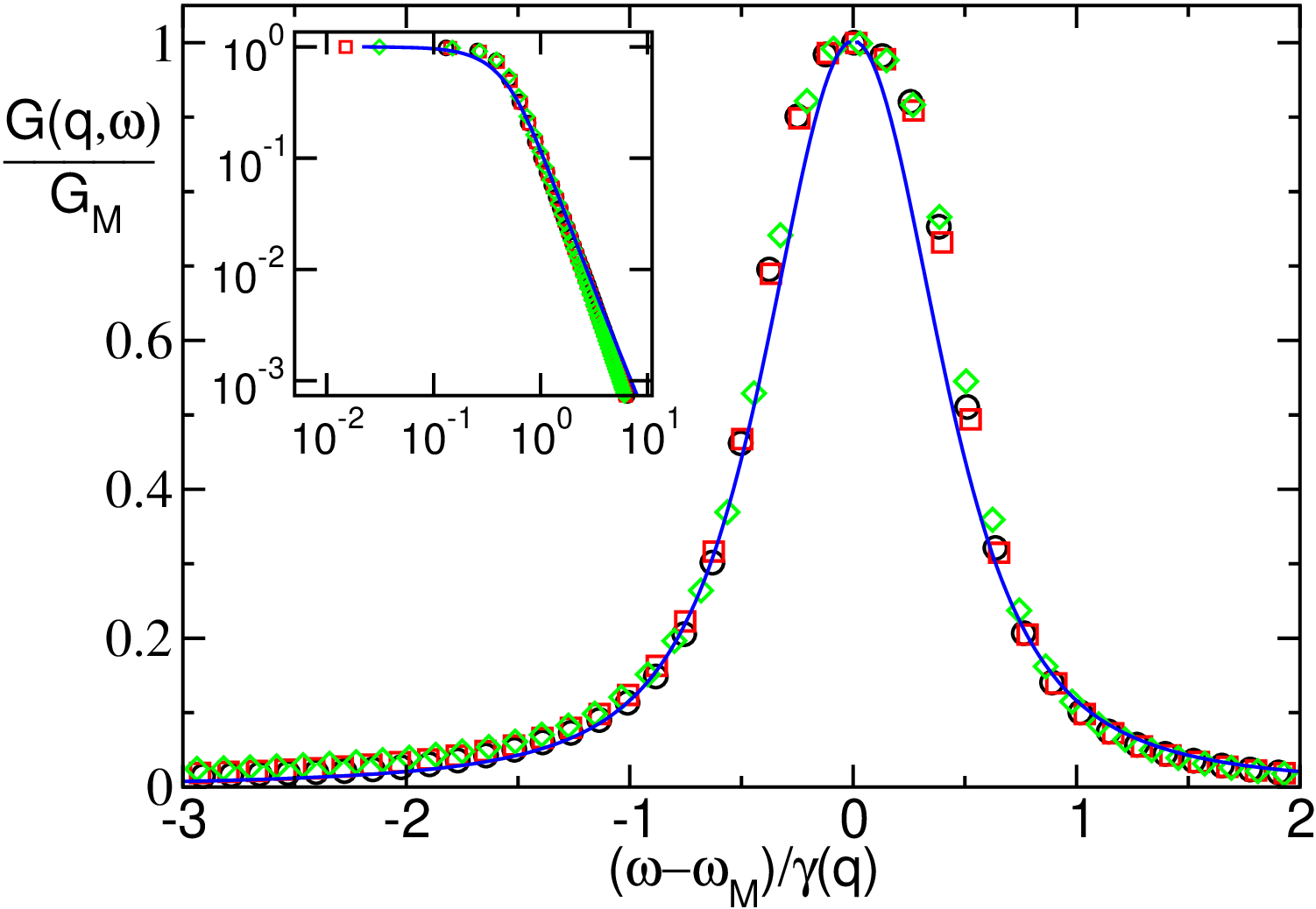}
\caption{Fourier transform $G(q,\omega)$ of the correlation functions for
$\varepsilon=1$, $N=2000$ and the three wavenumbers, $q=\pi/100$ (circles),
$2\pi/100$ (squares), and $3\pi/100$ (diamonds). The solid line corresponds to
the approximate analytic theory, i.e. it is obtained by Fourier
transforming the function $C(q,t)$ defined in (\ref{gf3},\ref{smallx}).
The same curves are plotted in the inset in log-log scales, where only positive
frequencies are shown.}
\label{spettri}
\end{center}
\end{figure}

To assess the validity of the above calculation, we have numerically integrated
equations (\ref{mct}) by the Euler method for the original dispersion relation
$\omega(q)$ and different values of $\varepsilon$. We have verified that a time
step $\Delta t = 0.01$ guarantees a good numerical accuracy over the explored
time range. The Fourier transform $G(q,\omega)$ is plotted in
figure \ref{spettri} for three different $q$-values versus $\omega -
\omega_{M}(q)$, where $\omega_{M}(q)$ is the frequency corresponding to the
maximal value $G_{M}$ of the spectrum (this is  equivalent to removing the
oscillating component from $G(q,t)$). Furthermore, in order to test relation
(\ref{gf3}), the vertical axis is scaled to the maximum $G$-value, while the
frequencies are divided by the half--width $\gamma(q)$ at half of the maximum
height. This latter quantity can be interpreted as the inverse lifetime of 
fluctuations of wavenumber $q$. The good data collapse confirms the existence of
a scaling regime. Moreover, the numerical data is compared with the approximate
analytical expression, obtained by Fourier transforming the function defined in
(\ref{gf3}) and (\ref{smallx}) (solid curve). The overall agreement is
excellent; there are only minor deviations for small values of $\omega
-\omega_M$  where (\ref{smallx}) is not strictly applicable. Moreover,
in the inset of figure \ref{spettri}, where the same curves are plotted using
doubly logarithmic scales, one can see that the lineshapes follow the predicted
power-law, $\omega^{-7/3}$, over a wide range of frequencies.  This scaling has
been deduced by Fourier transforming $g(x)$ in equation~(\ref{smallx}). 

\begin{figure}
\begin{center}
\includegraphics[clip,width=7.5cm]{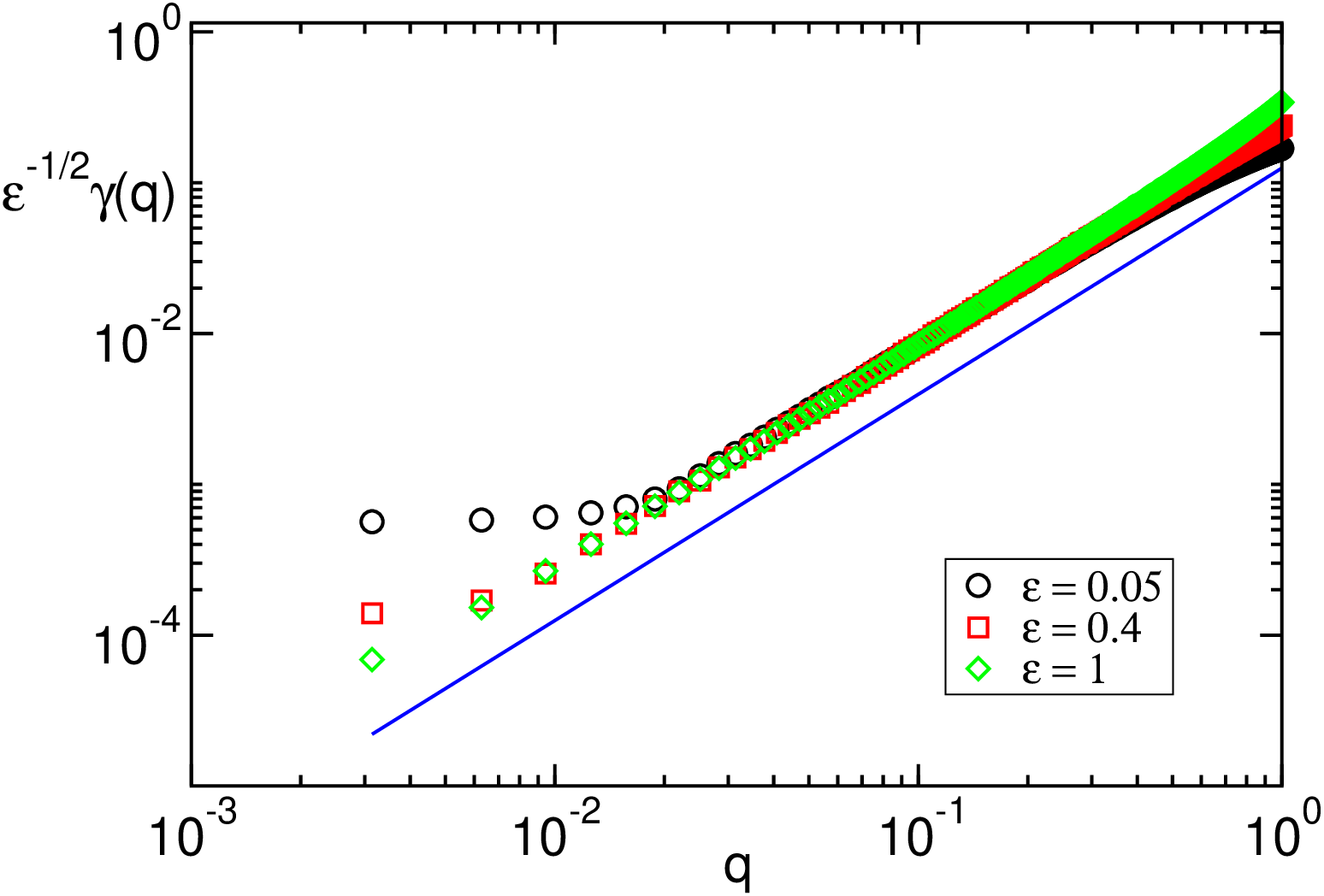}
\caption{Scaling of the linewidth $\gamma(q)$ of $G(q,\omega)$ with $q$ 
for three different values of the coupling constant $\varepsilon$ and $N=2000$.
The solid line corresponding to the expected power law $q^{3/2}$ is plotted for
reference.}
\label{gamma2}
\end{center}
\end{figure}

\begin{figure}
\begin{center}
\includegraphics[clip,width=7.5cm]{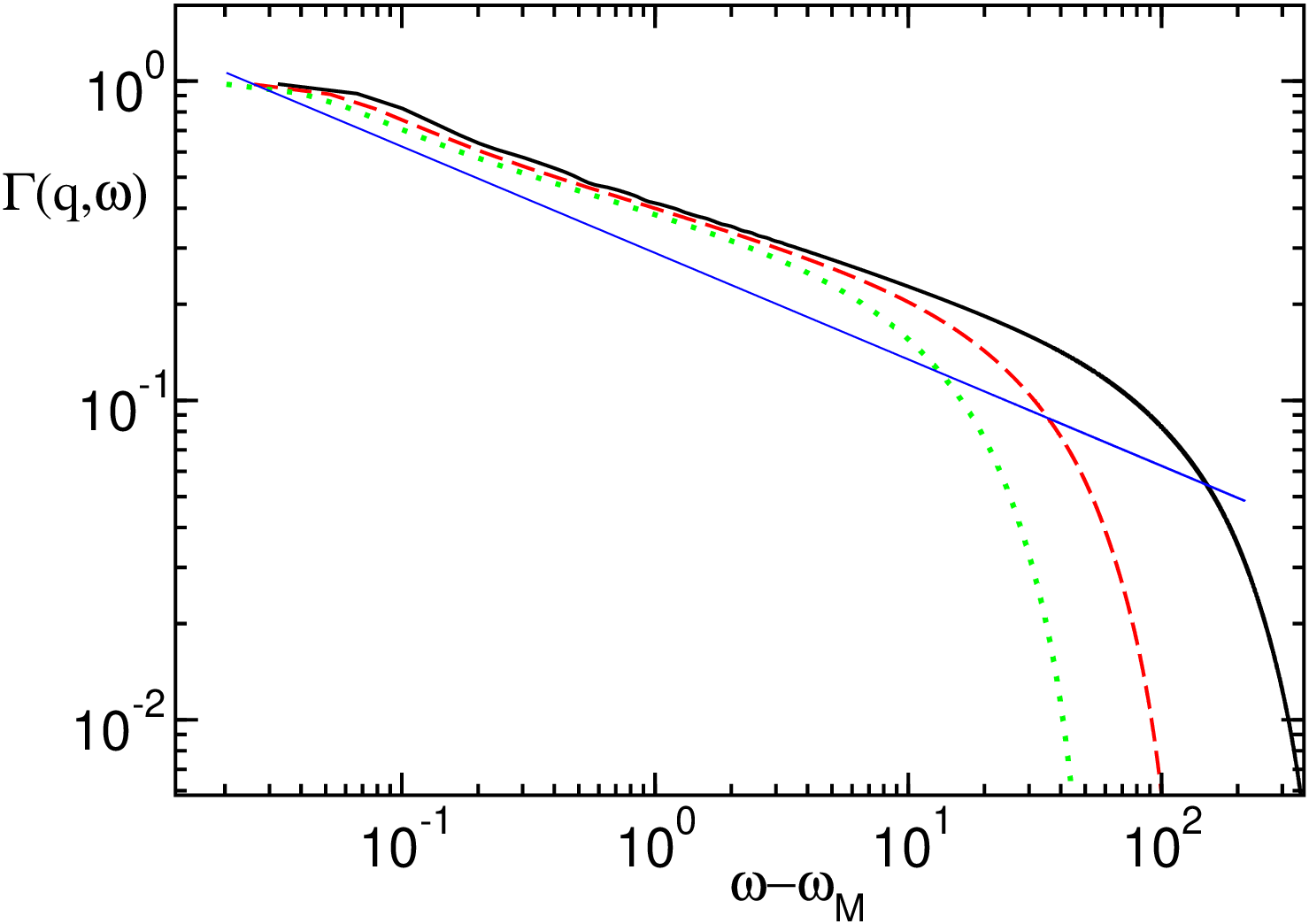}
\caption{ 
Fourier transform of the memory function $\Gamma(q,\omega)$ for the same
parameter values as in figure~\ref{spettri} and for three wavenumbers, namely, 
$\pi/100$ (solid line), $2\pi/100$ (dashed line), and $3\pi/100$ (dotted line).
The $\omega$ axis is suitably shifted in order to get rid of the oscillatory
peak at frequency $q$, so that $\Gamma(q,\omega)$ reduces to $M(q,\omega)$.
For  comparison, the expected power-law $\omega^{-1/3}$ is also presented
(thin solid line).}
\label{fig:deriv}
\end{center}
\end{figure}

In figure \ref{gamma2} we show that $\gamma(q)$ is indeed proportional to
$\sqrt{\varepsilon}\, q^{3/2}$. It is noteworthy to remark that the agreement is
very good also also for a relatively large value of the coupling constant such
as $\varepsilon = 1$, although the slowly varying envelope approximation is not
correct for large wavenumbers. The deviations observed at small $q$-values for
small couplings are due to the very slow convergence in time. Better
performances could be obtained by increasing both $N$ and the integration time
($10^4$, in our units), however, well beyond our current capabilities. Finally,
we have plotted the Fourier tranform of the memory function (see
figure \ref{fig:deriv}) to check its scaling behaviour. By Fourier transforming
equation~(\ref{m}), one finds a behaviour like $\omega^{-1/3}$ which is
reasonably well reproduced in figure \ref{fig:deriv}). Once again, in order to
obtain a larger scaling range, it would be necessary to consider significantly
larger system sizes.

Making use of equations ~(\ref{gf3},\ref{smallx}), it can be shown that the
memory function $\Gamma$ contains terms of the form $ q^2 \, e^{\pm
iqt}/t^{2/3}$, i.e. it oscillates with a power--law envelope. Accordingly, its
Laplace trasform has  branch-cut singularities of the form  $q^2/(z \pm q
)^{\frac{1}{3}}$. This finding is not consistent with the heuristic assumption
made in \cite{E91,L98} and the result of \cite{li}, where {\it simplified} MCT
equations were solved with the Ansatz $\Gamma(q,z)=q^2\mathcal{V}(z)$. In
addition, the numerical solution does not show any signature of the
$q^2/z^{1/3}$ dependence found in ~\cite{li}.  For instance, it would imply a
peak at $\omega=0$ in the spectrum of $\Gamma$ which is, instead, absent in our
data.

\section{Quartic nonlinearities}
\label{s:quartic}

In this section, we turn our attention to the case of a vanishing $g_3$.
This the case of symmetric potentials (with respect to the equilibrium
position), but it is more general than that.
The kernel of the mode coupling equations writes in this case
\begin{equation}
\Gamma(q,t)= \,\omega^{2}(q)
\,\Big(\frac{2 \pi}{N} \Big)^2 \sum_{p+p'+p''-q=0,\pm\pi}  
\,G(p,t) G(p',t) G(p'',t)  \quad .
\label{memquart}
\end{equation}
Equation (\ref{mct}) is still formally valid as it stands, with the 
(bare) coupling constant now given by 
$\varepsilon = {15 (g_4 \, k_BT  / 2\pi)^2}$ \cite{L98}. 

\subsection{Analytical results}

One can repeat the same analysis performed for the cubic case. By substituting
expression (\ref{g}) into (\ref{memquart}), we must now deal with
a double integral,
\begin{eqnarray}
\fl \Gamma(q,t) =  \omega^2(q)\int_{-\pi}^{\pi}dp
\int_{-\pi}^{\pi}dp'
\Big[ C(p,t)C(p',t)C(q-p-p',t)
e^{i[\omega(p)+\omega(p')+\omega(q-p-p')]t} \nonumber\\
  +C(p,t)C(p',t)C^{\ast}(q-p-p',t)
e^{i[\omega(p)+\omega(p')-\omega(q-p-p')]t} \nonumber \\
  +C(p,t)C^{\ast}(p',t)C(q-p-p',t)
e^{i[\omega(p)-\omega(p')+\omega(q-p-p')]t} \nonumber\\
  +C(p,t)C^{\ast}(p',t)C^{\ast}(q-p-p',t)
e^{i[\omega(p)-\omega(p')-\omega(q-p-p')]t} + c.c.\Big]\quad.
\end{eqnarray}

By neglecting the non-resonant terms in the limit $q \rightarrow 0$, the above
expression simplifies to
\begin{equation}
\fl \Gamma(q,t) = q^2 \int_{-\pi}^{\pi}dp
\int_{-\pi}^{\pi}dp' 
\Big[C^{\ast}(p,t)C^{\ast}(p',t)C(q-p-p',t)e^{i \omega'(p+p')qt} +c.c.\Big]
\end{equation}
which, by recalling that $\omega' \approx 1$, is formally equivalent to
equation~(\ref{Gamma}). Altogether, equation~(\ref{eqc}) applies also to the quartic
potential, the kernel $M$ being now given by
\begin{equation}
M(q,t) = q^2 \int_{-\pi}^{\pi}\, dp \int_{-\pi}^{\pi} \, dp' \, 
C^{\ast}(p,t)C^{\ast}(p,t) C(q-p-p',t)
\label{eqquart}
\end{equation}
Equations (\ref{eqc},\ref{eqquart}) satisfy the new scaling relations
\begin{equation}
C(q,t) \;=\; g(\sqrt{\varepsilon}t q^2) \quad, \quad
M(q,t) \;=\; q^{4}f(\sqrt{\varepsilon} t q^2)
\label{gf4}
\end{equation}
As in the cubic case, the functions $f$ and $g$ can be determined by 
substituting
the previous expressions into equations~(\ref{eqc},\ref{eqquart}). 
Upon setting $x=\sqrt{\varepsilon} q^2t$ one obtains,
\begin{eqnarray}
&&\frac{dg}{dx} = -\int_0^x dy \, f(x-y) g(y)
\label{gx4}\\
&&f(x) = x^{-1}\int_{-\infty}^{+\infty} dy \, \int_{-\infty}^{+\infty}dy'\,
g^{\ast}(y^2)g^{\ast}(y'^2)g(\mid x^{1/3}-y-y'\mid ^{2})   
\end{eqnarray}
In the limit $x \rightarrow 0$, one can analytically derive the 
asymptotic behaviour that is simply
\be
f(x)=\frac{a}{x}
\label{fx4}
\ee
where
\be
a=\int_{-\infty}^{+\infty}dy \, \int_{-\infty}^{+\infty}dy'\,g^{\ast}(y^2)
g^{\ast}(y'^2)g(\mid y+y'\mid ^{2})
\ee
which, in turn, implies that
\be
M(q,t)=\frac{a}{\varepsilon^{1/2}}\frac{q^2}{t}
\label{emme4}
\ee
By inserting equation~(\ref{fx4}) into (\ref{gx4}) one finds
\be
\frac{dg}{dx}=-\int_{0}^{x}dy \, 
a\frac{g(y)}{x-y}=-a\int_{0}^{x}dy\,g(y)\delta(x-y)=-ag(x)
\ee
whose, properly normalized solution, is
\be
g(x)=\frac{1}{2}\exp(-ax)
\label{ggg}
\ee
By assuming that (\ref{fx4}) is valid in the entire $x$-range, we find that
the Laplace transform of $C(q,t)$ is
\begin{equation}
C(q,z)=\frac{i}{iz+b q^2}
\label{cquart}
\end{equation}
where $b$ is a suitable constant. Remarkably, the correlation function is a
simple exponential, and its Fourier transform is a Lorentzian curve.
Note that also in this case we obtain the Mittag-Leffler function with index 
$\mu =1$, which is precisely an exponential \cite{maina}.

\subsection{Numerical analysis}

\begin{figure}[ht]
\begin{center}
\includegraphics[width=7.5cm]{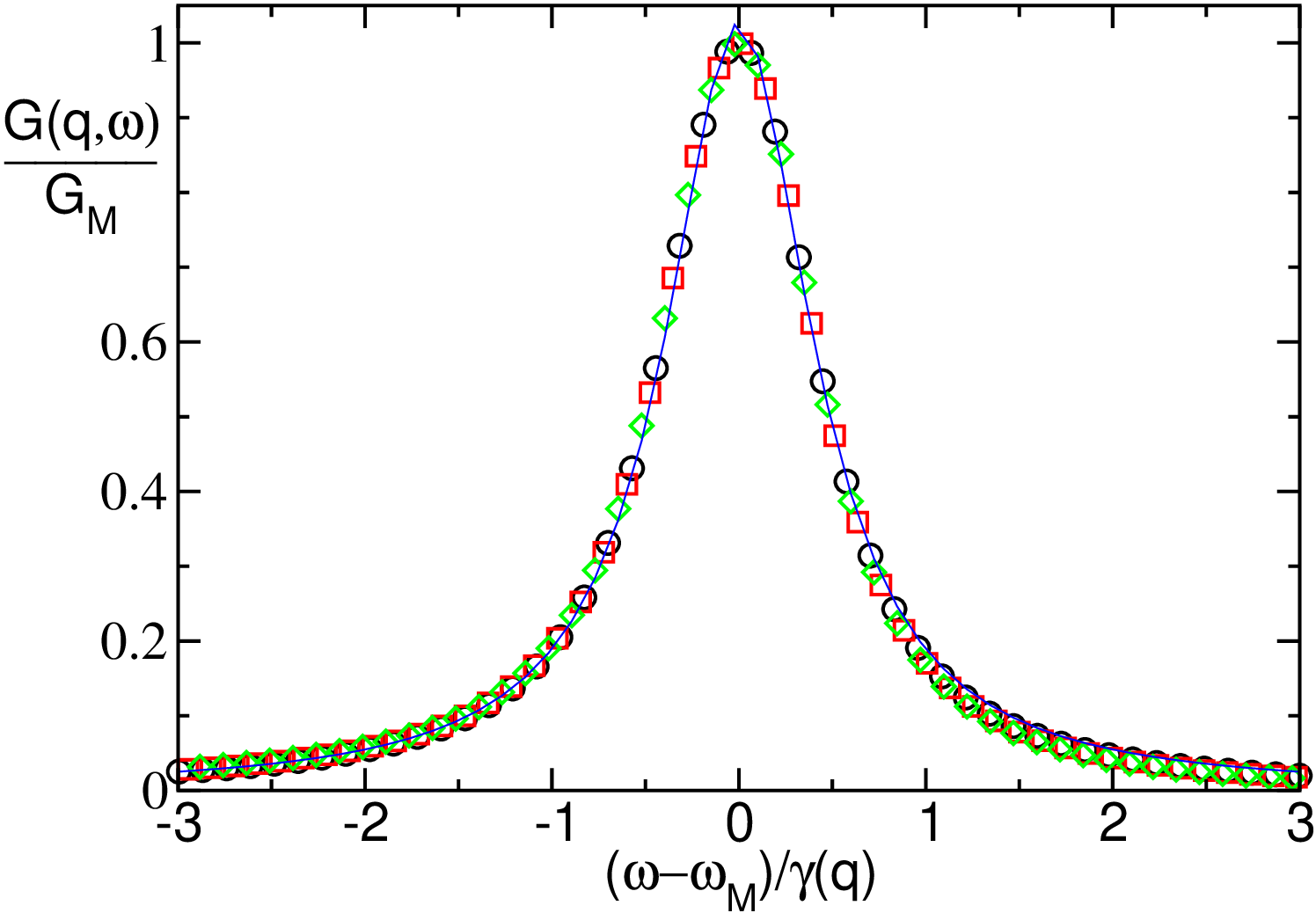}
\caption{Quartic case: Fourier transform $G(q,\omega)$ of the correlation function
for $\varepsilon=1$, $N=2000$ and the three wavenumbers $q=\pi/100$ (circles),
$2\pi/100$ (squares), and $3\pi/100$ (diamonds). The solid line corresponds
to the approximate analytical solution obtained by Fourier transforming
the expression defined in (\ref{gf4},\ref{ggg}).}
\label{spettri4}
\end{center}
\end{figure}

Also for the quartic case we have compared the analytical solutions with
the numerical integration of the full mode-coupling equations. In
figure \ref{spettri4} we have plotted the Fourier transform of the correlation
function $G(q,\omega)$, by following the same approach used in the cubic case,
for three different $q$-values. The $y$ axis is scaled by $G_{M}$ and the
frequencies are divided by half the maximum width $\gamma(q)$. The good data
collapse confirms the existence of a scaling regime (see (\ref{gf4})).
Moreover, at variance with the previous case, the approximate analytic solution
agrees very well with the numerical data in the entire range, including
the region around the maximum, where correctness is not a priori obvious.

\begin{figure}[ht]
\begin{center}
\includegraphics[width=7.5cm]{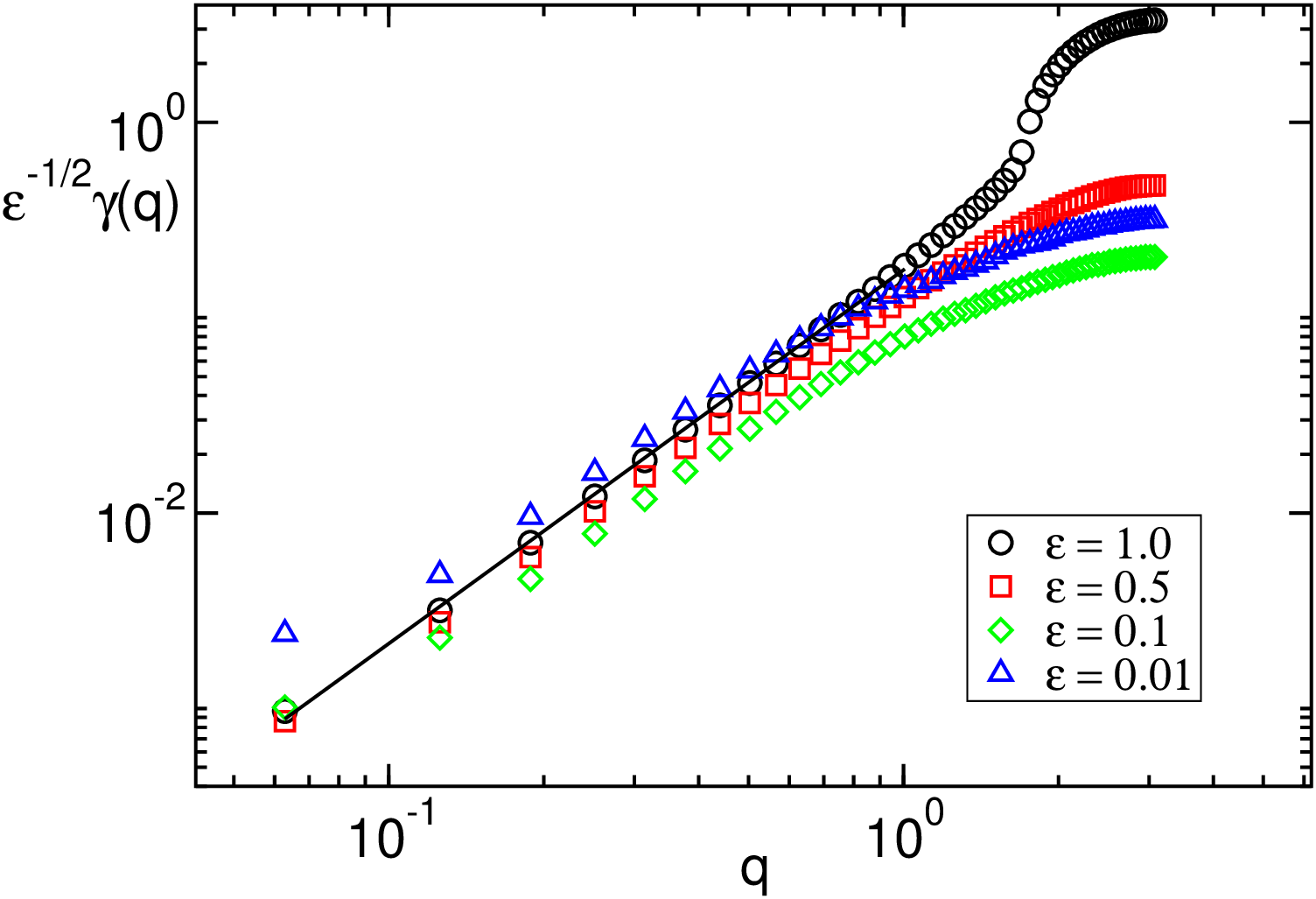}
\caption{Scaling of the linewidth $\gamma (q)$ of $G(q,\omega)$ for four different
values of the coupling constant $\varepsilon$ and $N=100$. The curves are obtained
by integrating over $10^4$ time units. The solid line with slope 1.91 is a fit
of the data corresponding to $\varepsilon=1$.}
\label{gamma}
\end{center}
\end{figure}

A further test of the theory can be made by determining the behavior of the
linewidths $\gamma(q)$, which our theory predicts to scale as
$\sqrt{\varepsilon}q^2$. In order to verify this scaling behaviour, it is
necessary to test small wavenumbers, but the smaller $q$, the longer it takes to
reach the asymptotic regime. Moreover, the rate of convergence is controlled
also by the coupling strength $\varepsilon$. In figure \ref{gamma}, we plot the
data corresponding to four different values of $\varepsilon$. They show that
already for $\varepsilon = 0.1$, the chosen time span ($10^4$ units) is not
enough to ensure a reasonable convergence for the smaller $q$-values.
Nevertheless, a fit of the tail of the curve corresponding to $\varepsilon=1$
reveals a reasonable agreement with the theoretical prediction (the slope is
indeed equal to 1.91). Unfortunately, choosing yet larger values of
$\varepsilon$ would descrease the width the scaling range.

\begin{figure}[ht]
\begin{center}
\includegraphics[width=7.5cm]{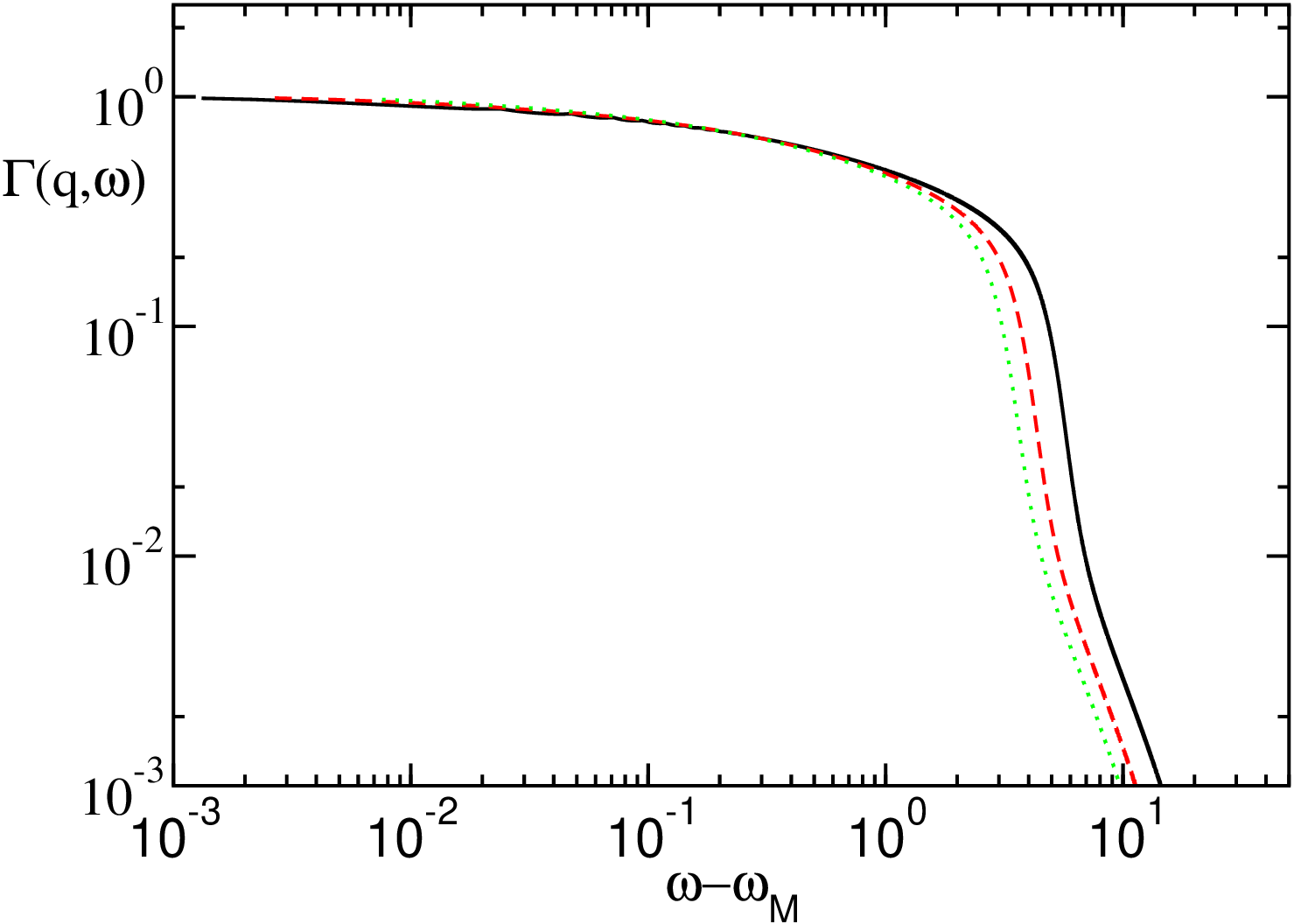}
\caption{Fourier transforms of the memory function for three values of 
the wave-vector for $q=\pi/100$ (solid line), $2\pi/100$ (dashed line), and 
$3\pi/100$ (dotted line). For small frequencies they tend to a constant 
as we expect by analytic solution.}
\label{mem}
\end{center}
\end{figure}

Finally, in figure \ref{mem} we plot $\Gamma(q,\omega)$  versus $\omega -
\omega_M$ . At variance with the cubic case, the spectrum does not exhibit
any low-frequency divergence. This is consistent with the analytic solution,
since the Fourier transform of equation~(\ref{emme4}) saturates for $\omega
\rightarrow 0$.

\section{Energy current correlator}
\label{s:cur}

In the previous sections we have solved the mode-coupling equations as it
applies to monoatomic chain of anharmonic oscillators for two distinct leading 
nonlinearities. Here, we use such results to determine the transport properties
of the underlying lattice. In particular, we are interested in the thermal
conductivity as defined by the Green-Kubo formula,
\be
\kappa\;\propto \; \int_{0}^{\infty}\, \langle  J(t)J(0) \rangle\,dt
\label{kubo}
\ee
where $J(t)$ is the total heat current, and $\langle \ldots \rangle$ brackets
denote the equilibrium average. A general expression for the heat current has
been derived in \cite{rep}. For the determination of the scaling behaviour it
is sufficient to consider only the harmonic part, which can be expressed as
a sum over wavevectors
\be
J\;=\; \sum_{q} b(q) Q(q,t)P^*(q,t) 
\ee
where $Q(q,t)$, and $P(q,t)=\dot Q(q,t)$ are the canonical variables in Fourier
space and 
\be
b(q)=i\omega(q)\frac{\partial \omega(q)}{\partial q}\quad.
\ee 
This amounts to disregarding higher-order
terms which are believed not to modify the leading behaviour. Under the same
approximations that allow deriving the mode-coupling equations, i.e. by
neglecting correlations of order higher than two, one obtains,
\be
\fl
\langle J(t)J(0)\rangle = \sum_{q}\mid b(q)\mid^2\{ \langle Q(q,t)Q(q,0)
\rangle \langle P(q,t)P^*(q,0)\rangle +\langle Q(q,t)P^*(q,0)\rangle^2 \}
\ee
This expression can be further simplified under the assumption
$P(q) \approx \omega(q)Q(q)$, that is certainly valid in the small $q$-limit.
Altogether, this leads to the expression proposed in \cite{li},
\be
\langle J(t)J(0) \rangle \; \propto \; 
\sum_{q} \, \Big(\frac{d\omega (q)}{dq}\Big)^2 G^2(q,t) 
\label{flusso}
\ee
By taking into account the general expression (\ref{g}), one obtains
\be
\fl
\langle J(t)J(0) \rangle \; \propto \; \int dq \,
\Big[ C^2(q,t)e^{2iqt}+{C^*}^2(q,t)e^{-2iqt}+2C(q,t)C^*(q,t) \Big]
\ee 
where, as usual in the thermodynamic limit, we have replaced the sum with
an integral and have made use of the fact that for small $q$ (the integral is
dominated by the small-$q$ terms), $\omega(q) \approx q$.  
We can finally use (\ref{gf3}) to determine the scaling behaviour of the
heat flux autocorrelation. Since the contribution of the terms proportional
to ${\rm e}^{2iqt}$ is negligible, due its ``rapid" oscillations, one obtains,
\be
\langle J(t)J(0)\rangle \;\propto \; \int dq \, g^2(\sqrt{\varepsilon} t q^{3/2})
\;\propto\; \frac{1}{t^{2/3}}
\ee
The resulting integral in equation~(\ref{kubo}) thereby diverges, revealing that
the effective conductivity of an infinite lattice is infinite. In order to
estimate the dependence of $\kappa$ on the system size $L$ it is customary to
restrict the integration range to times smaller than $L/c$ where $c$ is the
sound velocity. As a result, one finds that $\kappa(L)$ diverges as $L^{1/3}$,
i.e. $\alpha =1/3$. Therefore we conclude that mode coupling predictions are 
now in full agreement with those of the renormalization group \cite{NR02}.

\begin{figure}
\begin{center}
\includegraphics[width=7.5cm]{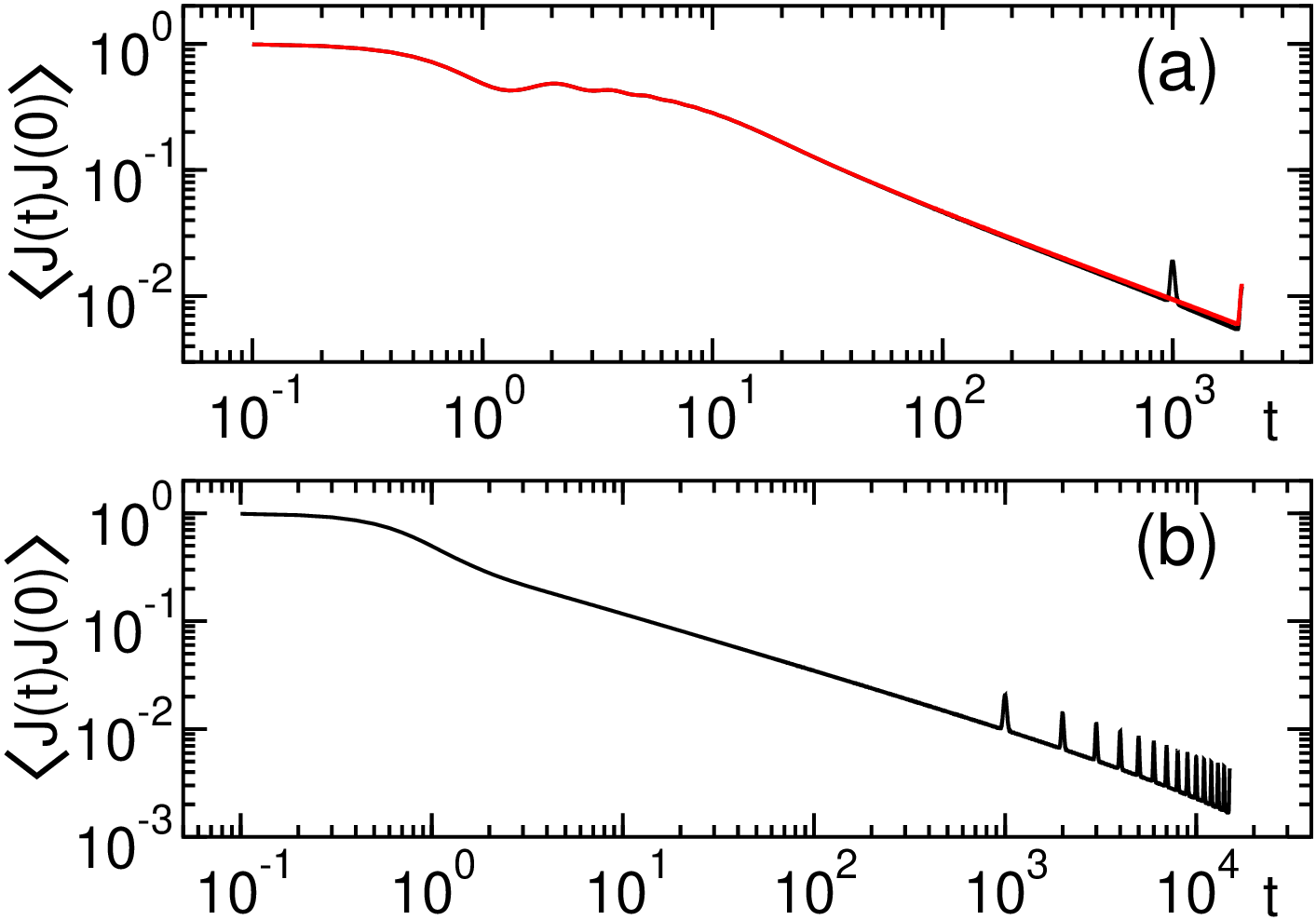}
\caption{Correlation function of the heat-flux for the cubic (a) and quartic (b)
case. The curves have been obtained for $N=2000$. In (a), also the outcome
of a calculation made with $N=4000$ are plotted to show the good overlap in this
time interval; the major difference is the obvious shift of the peak observed
for a time equal to the system length divided by the sound velocity.}
\label{corfl1}
\end{center}
\end{figure}

A similar analysis for the quartic case, leads, using equation~(\ref{gf4}), to
\be
\langle J(t)J(0) \rangle \;\propto \; \int dq \, g^2(q^{2}t)
\propto \frac{1}{t^{1/2}}
\ee
which, in turn, corrsponds to $\kappa \sim L^{1/2}$, i.e. $\alpha=1/2$.
Finally, we have compared the theoretical predictions with the computation of
the sum (\ref{flusso}) complemented by direct integration of the mode
coupling equations for $G(q,t)$ (see figure~\ref{corfl1}). A fit of the
scaling region yields a slope equal to $-0.69$ in (a) and $-0.55$ in (b) to be
compared with the theoretical expections that are respectively equal to $-2/3$
and $-1/2$. The deviations are analogous to those ones observed in the
previous explicit analysis of the behaviour of $G(q,t)$ itself and can thus
be attributed to finite size/time corrections.

\section{Conclusions}

In this paper we solved exactly the 1D MCT corresponding to a chain of nonlinear
oscillators. The only assumption we made is that the dynamics of both the
correlator $G(q,t)$ and the memory function $\Gamma(q,t)$ factorizes in the
product of a ``rapidly'' oscillating term times a slowly decaying function, see
equation~(\ref{g}). This Ansatz is confirmed \textit{a posteriori} by the
scaling of the analytic solution and by numerical calculations. The oscillating
behaviour of $\Gamma(q,t)$ represents an element of novelty that was disregarded
in previous studies \cite{L98,li} (see hovever the perturbative calculation in
\cite{L00}). In fact, taking the oscillations into account, we find that
the decay of the envelope $M(q,t)$ depends on the order of the leading
nonlinearity of the theory. In the cubic case, $M \propto t^{-2/3}$, while a
faster decay, $M \propto t^{-1}$, is found in the quartic one, see equations
(\ref{m},\ref{emme4}). As a consequence, in the two cases, the decay of  
$G(q,t)$ is, respectively, anomalous (i.e. nonexponential) and exponential, see
equations~(\ref{smallx},\ref{ggg}). Upon comparing with the solution of the
Markovian equation (\ref{markov}), we thus argue that the cubic case is
characterized by a diverging viscosity $\eta$, while a normal hydrodynamic
behaviour (finite $\eta$) is found in the quartic case. At the same time, the
other transport coefficient, the thermal conductivity $\kappa$, diverges in both
cases. Indeed, the energy--current autocorrelation (as estimated from
equation~(\ref{flusso})) shows a long--time tail, $t^{-2/3}$ and $t^{-1/2}$ in
the cubic an the quartic case, respectively. This implies that for large $L$,
$\kappa$ diverges as $L^{\alpha}$, with $\alpha=1/3$ (cubic nonlinearity) or
$\alpha=1/2$ (quartic nonlinearity). 

Since generically, the leading nonlinear term is cubic, we can conclude that our
analysis reconciles MCT with the renormalization-group prediction \cite{NR02}.
Accordingly, on the theoretical side there is no longer any doubt on the
existence of a wide universality class characterized by the exponent $\alpha
=1/3$.

The quartic case is less generic, since it requires the fine tuning of a
control parameter (as for phase transitions) or the presence of a
symmetry forbidding the existence of e.g. odd terms in the potential. This is,
for instance, the case of the FPU-$\beta$ model which, by construction, contains
only quadratic and quartic terms. In fact, the most accurate numerical findings
obtained for the FPU-$\beta$ model \cite{LLP03} are qualitatively compatible with
the scenario emerging from the analytic solution of the MCT in the quartic case,
since $\alpha >2/5$ and does not exhibit a clear saturation with the system
size.

The two-universality-class scenario emerging from our solution of the MCT is
consistent with that one outlined in \cite{canadesi}. There, the scaling
behavior has been estimated by complementing hydrodynamic and thermodynamic
arguments under suitable assumptions. Such an analysis, that applies to quartic
nonlinearity, predicts the same behavior for $\kappa$ as ours, as well as a
finite $\eta$. As the Authors of reference \cite{canadesi} pointed out, the
cubic case requires a fully self-consistent calculation which is precisely the
task we accomplished here. Moreover, as already remarked, we conclude that both
transport  coefficients diverge. In addition, we argue that since $M$ and
$\langle J(t)J(0) \rangle$ decay with the same power-law, $\eta$ should scale in
the same way as $\kappa$. A few words are also needed to comment about the
definition of the second class, which is identified in the present paper from
the vanishing of the cubic term in the expansion of the effective potential
(\ref{fpu}), while in \cite{canadesi} from the equality between
constant--pressure and constant--volume specific heats. At least in the
FPU-$\beta$ model both constraints are simultaneously fulfilled. More in
general, it seems plausible that these are two equivalent definitions of 
the same class but it is not certain.

Altogether, one of the important consequences of our analytical treatment is
that fluctuations and response of 1D models cannot be accounted for by simple
equations such as (\ref{markov}). Indeed, memory effects emerging from the
nonlinear interaction of long-wavelength modes are essential and cannot be
considered as small corrections. As a consequence, (\ref{markov})
should be replaced by equations like (\ref{fractional}), meaning that
hydrodynamic fluctuations must be described by generalized Langevin equation
with power-law memory \cite{wang,lutz}. This provides a sound basis for the
connection between anomalous transport and kinetics (superdiffusion in this
case) \cite{dku}.  

Finally, a few words on the consistency between this theoretical analysis and
the simulation of specific models. Over the years it has become clear that
numerical studies are affected by various types of finite-size corrections
which make it difficult if not impossible to draw convincing conclusions (and
this is one of the main reasons why we turned our attention to analytical
arguments). The ubiquity of slow corrections suggests to carefully consider all
the implicit assumptions made in the various theories, in the hope to
understand the possibly universal origin of such deviation or even (though
unlikely) to obtain corrections to the leading exponents. In this perspective,
we wish to point out that the MCT solved in this paper is based on the
assumption that the hydrodynamic properties are dominated by the coupling
among sound-modes. A more general and complete mode-coupling analysis should
consider also the coupling with thermal modes. General considerations indicate
that the asymptotic scaling properties are not affected by lower-order
corrections emerging in this more general approach. On the other hand,
numerical experiments seem to suggest that such corrections could affect
significantly finite size/time corrections to the scaling.

\section*{Acknowledgements}

We acknowledge useful discussions with H.~Van Beijeren. This work
is supported by the PRIN2005 project {\it Transport properties of classical  and
quantum systems} funded by MIUR-Italy. Part of the numerical calculations were
performed at CINECA supercomputing facility through the project entitled  
{\it Transport and fluctuations in low-dimensional systems}.

\end{document}